\documentclass[preprint,12pt]{elsarticle}

\usepackage{color}
\usepackage{amssymb}
\usepackage{epsfig}
\usepackage{epstopdf}
\usepackage{graphicx,color}
\usepackage[usenames,dvipsnames]{xcolor}
\usepackage[section]{placeins}
\usepackage{amsmath}
\usepackage{subfigure}
\usepackage[normalem]{ulem}

\usepackage{xcolor}

\colorlet{darkgreen}{green!50!black}
\colorlet{brightyellow}{yellow!75!red}
\colorlet{orange}{red!50!yellow}
\colorlet{darkblue}{blue!60!black}
\colorlet{darkred}{red!80!black}

\usepackage{soul}

\usepackage{mathtools}
\usepackage{mathrsfs}
\usepackage{bm}
\usepackage{relsize}
\usepackage{indentfirst}

\usepackage{xcolor}

\journal{Physics Letters B}
\begin{document}
\begin{frontmatter}
\title{Three-body bound states with zero-range interaction in the Bethe-Salpeter
approach}
\author{E. Ydrefors$^a$} 
\author{J. H. Alvarenga Nogueira$^a$}
\author{V. Gigante$^b$}
\author{T. Frederico$^a$}
\author{V.A. Karmanov$^c$}
\address{$^a$Instituto Tecnol\'ogico de Aeron\'autica, DCTA, 
12228-900 S\~ao Jos\'e dos Campos,~Brazil}
\address{$^b$Laborat\'orio de F\'\i sica Te\'orica e 
Computacional - LFTC,  Universidade Cruzeiro do Sul,\\  01506-000 S\~ao Paulo, Brazil} 
\address{$^c$Lebedev Physical Institute, Leninsky Prospekt 53, 119991 Moscow, Russia}

\date{\today}

\begin{abstract}
The Bethe-Salpeter equation for three bosons with zero-range interaction is solved for the first time. For comparison the light-front equation is also solved. 
The input is the two-body  scattering length and the outputs are the three-body binding energies, Bethe-Salpeter amplitudes and light-front wave functions. 
Three different  regimes are analyzed: ({\it i}) For weak enough two-body interaction the three-body system is unbound. ({\it ii}) For stronger two-body interaction a three-body bound state appears. It provides an interesting example of a deeply bound Borromean system. ({\it iii})  For  even stronger two-body interaction this state becomes unphysical with a negative mass squared. However, another physical (excited) state appears, found previously in  light-front calculations.  The Bethe-Salpeter approach implicitly incorporates  three-body forces of relativistic origin, which are attractive and increase the binding energy.
\end{abstract}
\end{frontmatter}

\section{Introduction}\label{intr}
The zero-range interaction model is, undoubtedly,  one of the oldest and most essential models in nuclear physics. 
It provides a reference framework and allows us to qualitatively grasp some important features of the nucleon-nucleon interaction. 
Understanding what happens in the limiting case of the zero-range interaction help us to clarify qualitatively the effect of a cutoff associated with a finite range interaction. That's why it is useful to find and compare the 
solutions 
for the zero-range interaction for different few-body systems in various approaches.  
It is  well known  that  in a non-relativistic
three-body  system with zero-range interaction 
the binding energy  is not limited from below (Thomas collapse \cite{thomas}). A variational proof of the Thomas effect in non-relativistic three-body systems can be found in \cite{CouJMP95} and  the non-relativistic limit of few-body systems 
for the $\lambda\varphi^4$ theory was
studied in  \cite{CouAP99}.
For the relativistic three-body   bound system with  zero-range interaction the covariant Bethe-Salpeter (BS) equation for the 
Faddeev component  was derived in \cite{tobias1}. In that work, the corresponding  three-body equation in the light-front  (LF) dynamics was also 
derived by projecting the BS equation on the LF plane. Later it was re-derived  independently of the BS approach,  
i.e., in the LF  framework only \cite{ck3b}.  In both papers~\cite{tobias1,ck3b} the LF equation was solved numerically.
In the aforementioned references it was concluded that the Thomas collapse is prevented in relativistic three-body systems, since the relativistic effects generates an effective repulsion at small distances.

The three-body BS equation with zero-range interaction was never solved so far. Finding its solution has remained for a long time 
an important and challenging problem. Of course, after avoiding the Thomas  collapse in the LF framework, used in the previous works where only the valence component was considered, one can hardly expect that 
it will again appear in the BS framework. However, there are other important questions which can be clarified 
by solving the three-body BS equation and comparing the result with the LF one. 
For example, higher Fock components can have a significant effect even in two-body systems as shown in~\cite{ffactor2016} and it is expected to be even more substantial in the three-body case \cite{km08}.

The aim of this paper is thus two-fold: 

({\it i}) We solve, for the first time, the three-body BS equation with the zero-range interaction. For this aim, we  transform the 
Minkowski   BS equation in a suitable form to be expressed
in the Euclidean space. From the point of view of time-ordered graphs appearing in  LF dynamics, the three-body BS equation takes into 
account extra graphs incorporating antiparticles. In fact, they generate the effective three-body forces of   relativistic origin. 
For the one boson exchange (OBE) interaction, the manifestation of the LF induced  three-body forces was  studied in \cite{km08}.
In the OBE model the three-body forces appear also in absence of antiparticles whereas, as it is found in that paper,  for the zero-range interaction, the intermediate 
antiparticles are mandatory for generating the three-body forces. Comparison of the results found by solving the LF and BS equations 
is instructive and it sheds light on the properties of the relativistic three-body systems with the zero-range interaction. 
We will calculate and compare also the dependencies of the LF and BS amplitudes on the transverse momenta.
Fully Poincar\'{e}-covariant computation of the nucleon's Faddeev amplitude  with a ladder dressed-gluon exchange interaction 
was performed in \cite{Eichmann1} (see \cite{Eichmann2} for a review).

({\it ii}) It turns out that though the three-body state studied in  \cite{tobias1,ck3b} for the interaction providing the existence of a 
two-body bound state is indeed the most low-lying physical state (with minimal {\em positive} three-body mass $M_3^2$), there exists another 
(non-physical) low-lying state with {\em negative} $M_3^2$. This is a ``heavy legacy" of the Thomas collapse.
Formally, from the point of view of the spectrum classification, 
the latter state is just the ground state (since it has the smallest $M_3^2$), whereas the state found in  \cite{tobias1,ck3b} (and interpreted as the ground state)  is the first 
excited state. By varying the two-body  scattering length, we can push the 
ground state into the domain of the positive mass $M_3^2$, so it becomes a physical state. In this situation the excited 
state found in \cite{tobias1,ck3b} does not exist anymore,  since 
 it was already driven into the continuous spectrum. This happens in both approaches -- the LF and BS ones, and 
the difference between them is in the numerical values of the parameters, as we will show here.  Below we will 
discover and study this true low-lying state. This is another aim of our work.

The rest of this paper is organized as follows.  In Sections \ref{bseq} and \ref{lfeq} we present the three-body BS and LF equations. In 
Sec. \ref{trans} is devoted to the derivation of the $k_{\perp}$-dependent amplitudes in terms of 
the LF wave function and the BS Euclidean amplitude. In Sec. \ref{levels} we compute the positions of the ground and first excited levels 
and study how they move depending on variation of the two-body interaction. Sec. \ref{LFBS} presents the numerical results 
for the LF wave function, BS amplitude and corresponding $k_{\perp}$-dependent amplitudes. Finally, in 
Sec. \ref{concl} we draw the conclusions.


\section{Bethe-Salpeter equation}
\label{bseq}
The zero-range three-body BS equation for the vertex function $v(q,p)$ from which the external propagators are excluded, for zero-range interaction, has the form \cite{tobias1}:
\begin{equation}\label{3b}
v(q,p)= 2i F(M_{12}) \int \frac {d^4k}{(2\pi)^4} \frac {i}{[k^2-m^2+i\epsilon]}
\frac {i}{[(p-q-k)^2-m^2+i\epsilon]} v(k,p).
\end{equation}
Here $v(q,p)$ is the Faddeev component and, besides the total momentum $p$, it depends on one four-momentum $q$ only. The function $F(M_{12})$ is the two-body zero-range scattering amplitude found in a relativistic framework.
It is given in \cite{tobias1,ck3b}. For completeness we cite it here, however, using as a parameter, the scattering length $a$:
\begin{equation}\label{F}
F(M_{12})=
\left\{
\begin{array}{cll}
\frac{\displaystyle{8\pi^2}}{\frac{\displaystyle{1}}
{\displaystyle{2y'_{M_{12}}}}\displaystyle{\log \frac{1+y'_{M_{12}}}
{\displaystyle{1-y'_{M_{12}}}}}
-\displaystyle{\frac{\pi}{2a m}}},&   \mbox{if $M_{12}^2<0$}
\\
&
\\
\frac{\displaystyle{8\pi^2}}{\frac{\displaystyle{\arctan y_{M_{12}}}}
{\displaystyle{y_{M_{12}}}}
-\displaystyle{\frac{\pi}{2am}}},& \mbox{if $0\leq M_{12}^2<4m^2$}
\end{array}\right.
\end{equation}
Its argument $M_{12}$ is two-body effective mass: $M_{12}^2=(p-q)^2$ and
$y'_{M_{12}}=\frac{\sqrt{-M_{12}^2}}{\sqrt{4m^2-M_{12}^2}}$,
$ y_{M_{12}}=\frac{M_{12}}{\sqrt{4m^2-M_{12}^2}}$.
%
If the two-body system has a bound state with the mass $M_{2}$,
then $a$ is positive and it is related to the bound state mass $M_{2}$ as: 
\begin{equation}\label{a}
a=\frac{\pi y_{M_{2}}}{2m\arctan y_{M_{2}}},\quad y_{M_{2}}=\frac{M_{2}}{\sqrt{4m^2-M_{2}^2}}.
\end{equation}
If $a<0$, the amplitude $F(M_{12})$ has no pole in the physical domain $0\leq M_{12} \leq 2m$, that is, the two-body bound state is absent. However, as we will see below, the three-body system still can be bound as a Borromean state.

As mentioned, to simplify finding the solution of eq. (\ref{3b}), instead of the Minkowski space BS equation (\ref{3b}), we will solve the corresponding 
integral equation in the Euclidean space. It provides the same spectrum, but  different amplitudes.

 The Euclidean equation is obtained by the Wick rotation of the integration contour, when it is possible. 
In Eq. (\ref{3b}) it is impossible: one can easily check that the position of singularities in the variable $k_0$ of the integrand in (\ref{3b}) prevents from this rotation. That is, the rotating contour crosses the singularities of the integrand. This was the obstacle in finding solution of Eq. (\ref{3b}). However, the shift of the rotation point changes the relative position of the rotating contour and singularities and might allow to avoid their crossings. We notice that the Wick rotation becomes possible after the following shift of variables:
\begin{equation}\label{var}
k=k'+\frac{1}{3}p,\quad q=q'+\frac{1}{3}p.
\end{equation}
After introducing  new functions:
$$
\tilde{v}(q',p)=v\left(q'+\frac{1}{3}p,p\right),\quad  \tilde{v}(k',p)=v\left(k'+\frac{1}{3}p,p\right)
$$
the equation (\ref{3b}) obtains the form:
\begin{equation}\label{3bp}
\tilde{v}(q',p)= 2iF({M'}^2_{12}) \int \frac {d^4k'}{(2\pi)^4} \frac {i^2 \tilde{v}(k',p)}
{\left[\left(k'+\frac{1}{3}p\right)^2-m^2+i\epsilon\right]\left[\left(\frac{1}{3}p-q'-k'\right)^2-m^2+i\epsilon\right]},
\end{equation}
where ${M'}^2_{12}=(\frac{2}{3}p-q')^2$.
In the three-particle rest frame, for example, the position of the pole (above the real axes) of the second propagator in (\ref{3bp}) 
in the variable $k'_0$ is in the point
$k'_0=k'_{01}$, where
\begin{equation}\label{pp1}
k'_{01}=  \eta' +i\epsilon-q'_0,\quad \eta'=\frac{1}{3}M_3-\sqrt{(\vec{k}+\vec{q})^2+m^2}.
\end{equation}
For the bound state $M_3<3m$, the value of $\eta'$ is always negative: $\eta'<0$. We rotate the line of integration over $k'_{10}$ by the angle $\phi$ and simultaneously replace $q'_0\to q'_0\exp(i\phi)$.  Then the pole and the contour move so that the pole never crosses the contour. 

The amplitude $F({M'}^2_{12})$  has also a pole at ${M'}^2_{12}=M_2^2-i\epsilon$, corresponding to the two-body bound state, if any. It generates two poles in $\tilde{v}(k',p)$ vs. $k'_0$. One can easily check that if $\frac{2}{3}M_3<M_2$ (this is the case, since the three-body binding energy per particle is larger that the two-body one), then these poles also do not prevent the Wick rotation. Therefore we can safely make the Wick rotation in Eq. (\ref{3bp}), in contrast to the Eq. (\ref{3b}).

In the rest frame, after Wick rotation by the angle $\phi=\pi/2$: $k_0=ik_4,\,q_0=iq_4$, and after integrating in (\ref{3bp}) over the angles between $\vec{k}$ and $\vec{q}$, we obtain the equation: 
\begin{equation}\label{3bE}
v_E(q_4,q_v)= 2 F(-{M'}^2_{12}) \int_{-\infty}^{\infty}dk_4\int_0^{\infty} \frac{dk_v}{(2\pi)^3}
 \frac {\Pi(q_4,q_v,k_4,k_v)}{\left(k_4-\frac{i}{3}M_3\right)^2+k_v^2+m^2}v_E(k_4,k_v),
\end{equation}
 where $q_v=|\vec{q}|$  (similarly for $k_v$),    
\begin{equation}
\Pi(q_4,q_v,k_4,k_v)=\frac{k_v}{2 q_v}\log\frac{\left(k_4+q_4+\frac{i}{3}M_3\right)^2+(q_v+k_v)^2+m^2}{\left(k_4+q_4+\frac{i}{3}M_3\right)^2+(q_v-k_v)^2+m^2}
\label{3bE5}
\end{equation}
and  ${M'}_{12}^2=(\frac{2}{3}iM_3+q_4)^2+q^2_v$. Namely Eq. (\ref{3bE}) will be solved numerically below.

By performing the complex conjugation in (\ref{3bE}) and (\ref{3bE}) and changing $k_4\to -k_4,\;q_4\to -q_4$, we find the same equation for 
$v^*_E(-q_4,q_v)$. Hence, $v^*_E(-q_4,q_v)=v_E(q_4,q_v)$, or:
\begin{equation}
{\rm Re}[v_E(-q_4,q_v)]={\rm Re}[v_E(q_4,q_v)],\quad {\rm Im}[v_E(-q_4,q_v)]=-{\rm Im}[v_E(q_4,q_v)].
\label{sym}
\end{equation}

\section{Light-front equation}
\label{lfeq}
In the LF framework, the equation for the Faddeev component of the three-body vertex function $\Gamma(k_{\perp},x)$
reads \cite{tobias1,ck3b}:\footnote{Eq. (11) from \cite{tobias1} differs from (\ref{eq30}) by the integration limits 
incorporating cutoffs which are absent in (\ref{eq30}).}
\begin{equation}\label{eq30}
\Gamma(k_{\perp},x)
=F(M_{12})\frac{\displaystyle{1}}{\displaystyle{(2\pi)^3}}
\displaystyle{\int_0^{1-x}}
\frac{\displaystyle{dx'}}{\displaystyle{x'(1-x-x')}}\;
\displaystyle{\int_0^{\infty}}
\frac{\displaystyle{d^2k'_{\perp}}}{\displaystyle{M_0^2-M_3^2}}\;
\Gamma\left(k'_{\perp},x'\right),
\end{equation}
where $M_0^2$ is the invariant mass squared of the intermediate three-body state:
\begin{equation}\label{M0}
M_0^2=\frac{\vec{k'}^2_{\perp}+m^2}{x'}+\frac{\vec{k}^2_{\perp}+m^2}{x}
+\frac{(\vec{k'}_{\perp}+\vec{k}_{\perp})^2+m^2}{1-x-x'}.
\end{equation}
The two-body scattering amplitude $F(M_{12})$ is still given by Eq. (\ref{F}), but its argument $M_{12}$ -- two-body effective mass  -- is defined now as 
$$
M^2_{12}=(1-x)M_3^2-\frac{k_{\perp}^2+(1-x)m^2}{x}.
$$

In general, the Faddeev component depends on all the variables $\vec{k}_{1,2,3\perp},x_{1,2,3}$ constrained by the conservation laws:
 $\vec{k}_{1\perp}+\vec{k}_{2\perp}+\vec{k}_{3\perp}=0,\; x_1+x_2+x_3=1$.
Due to the zero-range interaction, $\Gamma(k_{\perp},x)$ depends on one pair of these variables only \cite{tobias1} which we denote 
$\vec{k}_{\perp}$ and $x$. 

\begin{figure}[ht!]
\begin{center}
\epsfig{figure=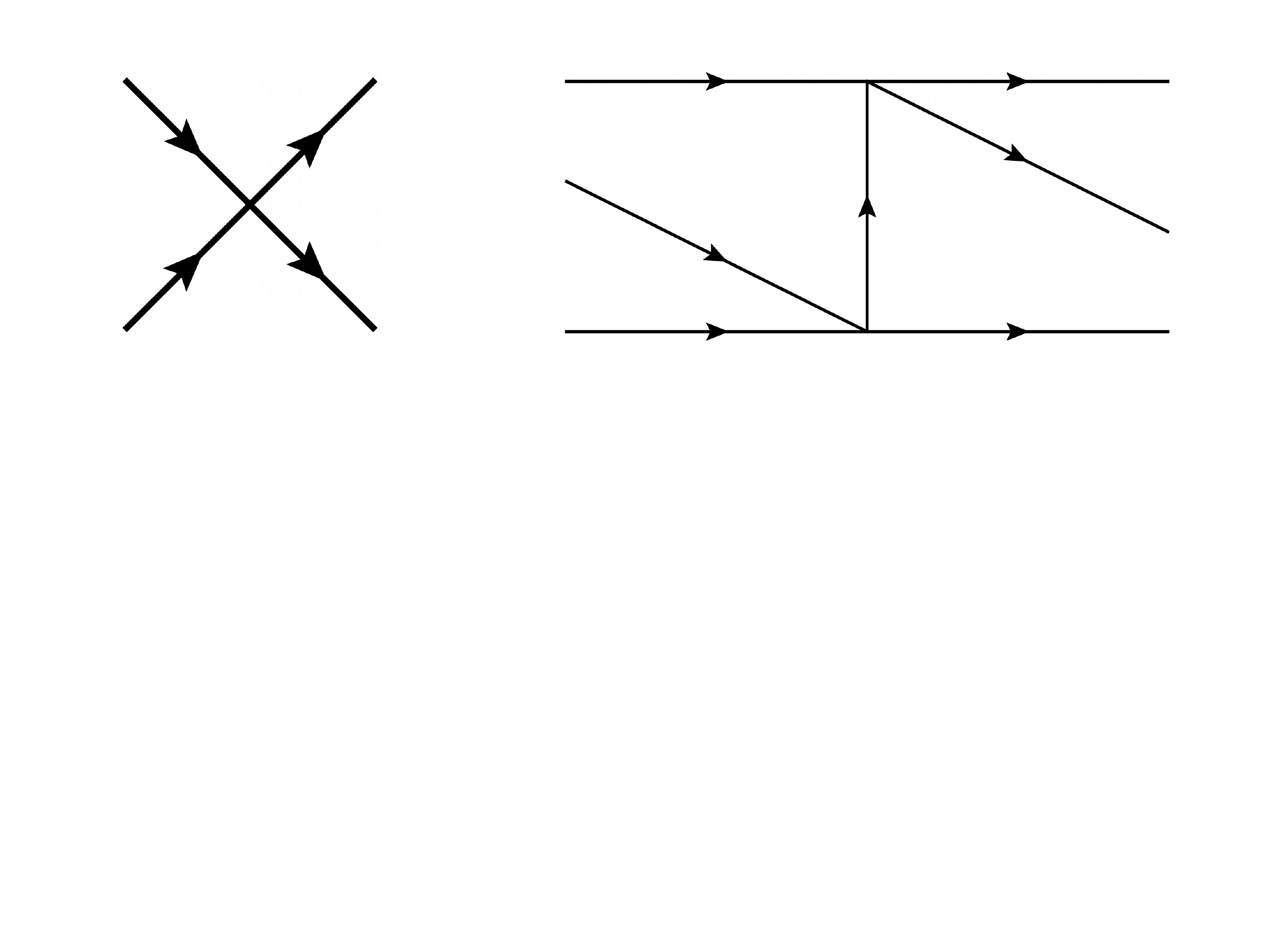,width=12cm}
\end{center}
\vspace{-6cm}
\caption{The elementary two-body cross graph $2\to 2$ is shown in the left panel, from which all the Feynman graphs for  the zero-interaction are composed.  The graph for the lowest order Feynman three-body amplitude $3\to 3$, composed by two elementary cross graphs (left panel)  is shown in the right panel. } 
\label{Fig1}
\end{figure}
\begin{figure}[!hbt]
\centering
\epsfig{figure=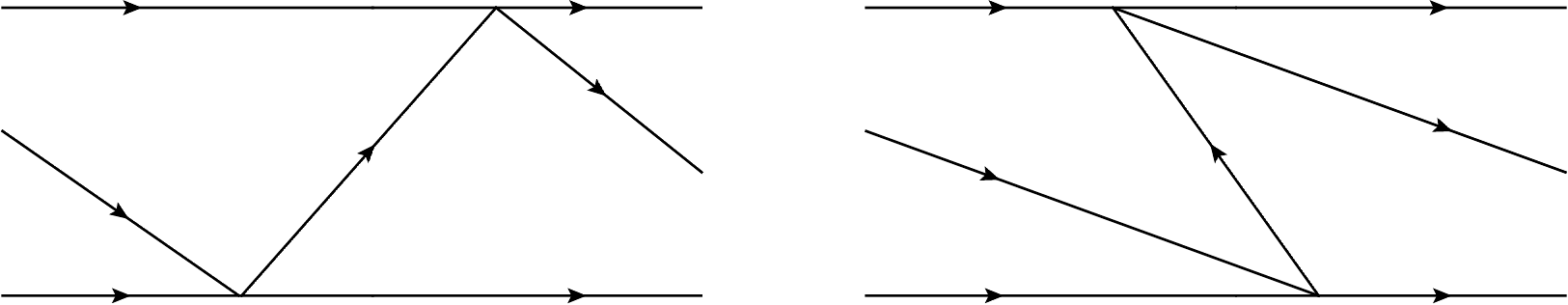,width=12cm}
\caption{The three-body LF graphs obtained by time-ordering of the Feynman graph shown in right panel of Fig. \protect{\ref{Fig1}}.}\label{Fig2} 
\end{figure}
\begin{figure}[!hbt]
\centering
\epsfig{figure=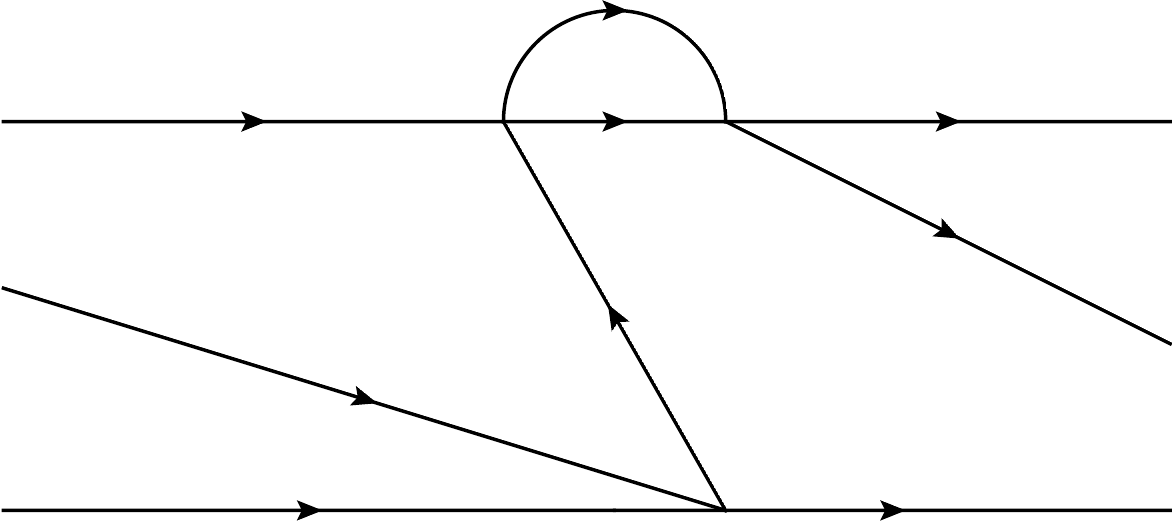,width=6.5cm}\qquad
\epsfig{figure=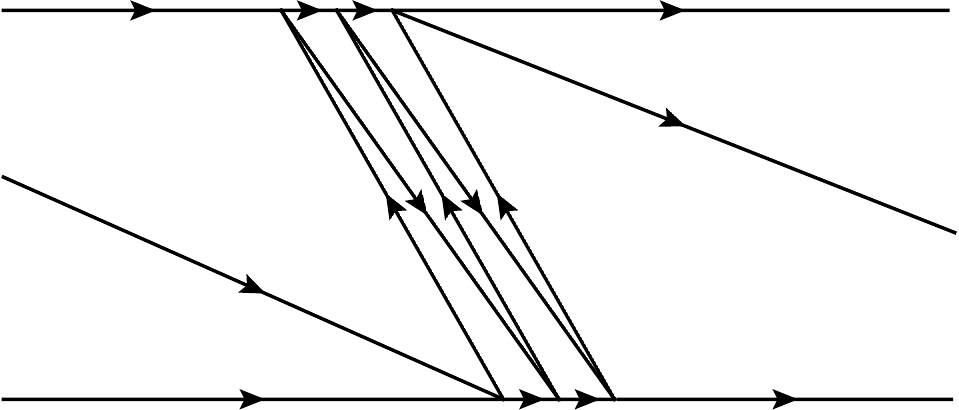,width=6.5cm}
\caption{Examples of many-body intermediate state  contributions to the LF three-body forces.}\label{Fig3} 
\end{figure}

Though both Eqs.  (\ref{3b}) and (\ref{eq30}) correspond to the zero-range interaction, the underlying dynamics is not identical. It is instructive to discuss this difference.
The cross graph shown in left panel of Fig.~\ref{Fig1} is the elementary two-body Feynman graph --  a ``building block", associated with the
two-body interaction, from which, connected by the propagators, all the graphs 
contributing to Eq. (\ref{3b}) are constructed. 
The lowest order three-body Feynman graph, constructed from  two of these two-body interaction ``blocks", 
is shown in the right panel of Fig.~\ref{Fig1}. 
The  LF graphs are obtained from the Feynman ones  by the time-ordering of all the vertices (omitting the graphs with creation from vacuum,
 when they appear).  One Feynman graph  in the right panel of Fig.~\ref{Fig1} (contributing, as the second iteration, in the BS equation (\ref{bseq})) corresponds to two time-ordered graphs shown 
 in Fig. \ref{Fig2}. For the ladder kernel, only the first graph of Fig. \ref{Fig2} contributes in the LF equation (\ref{lfeq}).  The graph $3\to 3$ shown in the right panel of 
Fig.~\ref{Fig2} contains the five-body intermediate state (including one antiparticle), but not the three-body one. 
This irreducible graph contributes to the effective three-body forces of relativistic origin.
The graphs of this type (or more complicated ones)  do not appear within the LF graph technique when one iterates the graph shown in left panel of Fig.~\ref{Fig2}. This is just the main difference between Eqs. (\ref{3b}) and (\ref{eq30}):  Eq.~(\ref{eq30}) contains the 
three-body intermediate states only (like Fig.~\ref{Fig2}, left panel), whereas Eq.~(\ref{3b}), from the point of view of LF dynamics, 
takes into account the graphs of the type shown in the  right panel of Fig.~\ref{Fig2},  and also more complicated ones.
 Examples of more complicated LF graphs are shown in Fig.~\ref{Fig3}. They provide  further contributions to the three-body forces of 
 relativistic origin. They are not included in Eq. (\ref{lfeq}), in the ladder approximation. However, they implicitly are taken into account by the BS equation (\ref{bseq}) and they appear explicitly after its LF projection. Thus, the graph shown in the right panel of Fig.~\ref{Fig3} contains  up to nine particles in the 
 intermediate state. These examples show that the number of particles in the intermediate states is not restricted - it can be arbitrary.
By comparing the results found by Eqs. (\ref{3b}) and (\ref{eq30}), we study the influence on the binding energy coming from these graphs 
  with the anti-particles in the intermediate states, which are generating the LF effective three-body forces. We emphasize that  the graph shown in the right panel of Fig.~\ref{Fig1} does not generate the three-body forces in the BS equation (\ref{3b}).  The graphs shown in Fig.~\ref{Fig2} (right panel) and Fig.~\ref{Fig3} represent the three-body forces in the LF equation (\ref{eq30}). Note also that this model does not correspond to  the $\lambda\varphi^4$ theory since all the graphs incorporated by the interaction
in Eq.~(\ref{3b}) is still a small part of the full set of possible contributions to the
 kernel in the $\lambda\varphi^4$ theory.

\section{Amplitudes depending on transverse  momenta}\label{trans}
In the BS amplitude,  instead of $k=(k_0,\vec{k})$ one  can introduce the LF variables $k=(k^-,k^+,\vec{k}_{\perp})$, where $k^{\pm}=k_0\pm k_z$. 
The LF wave function is related to the integral over $k^-$ of the Minkowski BS amplitude.  There is no corresponding relation in the Euclidean space.  
However, since
the double integrals of the Minkowski BS amplitude over $k^-$ and $k^+$,  and of the Euclidean one over $k_4,k_z$  are the same (up to a Jacobian), this 
allows us to relate the integral of the Euclidean BS amplitude with an integral of the  LF wave function. 
The relation was found in \cite{SalPRC00} and used in \cite{cristian} for two-body systems.
Below we will derive this relation for the three-body case.

The full vertex function is given by the sum of the Faddeev components. Correspondingly,  the wave function is obtained from the vertex one by dividing by the 
energy denominator:
\begin{equation}\label{Psi}
\psi(\vec{k}_{1\perp},\vec{k}_{2\perp},\vec{k}_{3\perp},x_1,x_2,x_3)=\frac{\Gamma(\vec{k}_{1\perp},x_1)+\Gamma(\vec{k}_{2\perp},x_2)+\Gamma(\vec{k}_{3\perp},x_3)}
{M_0^2-M_3^2},
\end{equation}
where $M_0^2$  is defined by (\ref{M0}).

For the BS amplitude, the energy denominator is replaced by the product of three propagators:
\begin{equation}\label{Phiv}
i\Phi_M(k_1,k_2,k_3;p)=i^3\frac{v_M(k_1)+v_M(k_2)+v_M(k_3)}{(k_1^2-m^2+i\epsilon)(k_2^2-m^2+i\epsilon)(k_3^2-m^2+i\epsilon)},
\end{equation}
where $k_1+k_2+k_3=p$.

In the three-body case we start with a 4D integral (over $k_{14},k_{1z},k_{24}$ and $k_{2z}$) from the Euclidean BS amplitude.
We can integrate analytically over two of the four variables, which do not enter in the argument of $v$.  Similarly, we obtain a 2D integral (over $x_{1}$ and $x_{2}$) from the LF wave function. We can integrate analytically over one of these variables. 
In this way, for the LF wave function contribution we find:
\begin{eqnarray}\label{R}
 A^{LF}(\vec{k}_{1\perp},\vec{k}_{2\perp})= A^{LF}_1+ A^{LF}_2+ A^{LF}_3,\quad
A^{LF}_i=\int_0^1dx_1\; \Gamma(\vec{k}_{i\perp},x_1)\,\eta(\vec{k}_{1\perp},\vec{k}_{2\perp};x_1),
\end{eqnarray}
where 
$$
\eta(\vec{k}_{1\perp},\vec{k}_{2\perp};x_1)=-\frac{1}{2}\sqrt{\frac{\pi}{2}}\int_0^{1-x_1}\frac{dx_2}{a'x_2^2+b'x_2+c'},
$$
and $a'=E_{1\perp}^2-x_1 M_3^2,\; b'=-(1-x_1)E_{1\perp}^2+x_1[E_{2\perp}^2-E_{3\perp}^2+(1-x_1)M_3^2],\;
c'=E_{2\perp}^2-x_1 M_3^2.
$

For the Euclidean BS contribution we get: 
\begin{equation}\label{L}
A^{BS}(\vec{k}_{1\perp},\vec{k}_{2\perp})=A^{BS}_1+A^{BS}_2+A^{BS}_3,\,
A^{BS}_1=
\int \tilde{v}_E(k_{14},k_{1v})\,\beta(k_{14},k_{1z};\vec{k}_{1\perp},\vec{k}_{2\perp}) dk_{14}dk_{1z},
\end{equation}
$$
\beta(k_{14},k_{1z};\vec{k}_{1\perp},\vec{k}_{2\perp})=-\frac{\chi(k_{14},k_{1z};E_{2\perp},E_{3\perp})}
{\left[\left(k_{14}-\frac{i}{3}M_3\right)^2+k_{1z}^2+E_{1\perp}^2\right]},\;
\chi(k_{14},k_{1z};\vec{k}_{1\perp},\vec{k}_{2\perp}) = \int_0^1\frac{\pi dy}{ay^2+by+c},
$$
where $k_{1v}=\sqrt{k_{1z}^2+k_{1\perp}^2},\;E_{i\perp}=\sqrt{m^2+k_{i\perp}^2},\;  \vec{k}_{3\perp}=-(\vec{k}_{1\perp}+\vec{k}_{2\perp})$
and $$a=-k_{1z}^2-\left(k_{14}+i\frac{2}{3}M_3\right)^2,\; b=k_{1z}^2+\left(k_{14}+i\frac{2}{3}M_3\right)^2+E_{2\perp}^2-E_{3\perp}^2,\;
c=E_{3\perp}^2.$$
Analogous formulas are easily found for $A_2^{BS}$ and $A_3^{BS}$.

If the LF wave function is obtained by the LF projection of the BS amplitude, then $A^{LF}(\vec{k}_{1\perp},\vec{k}_{2\perp})$, 
Eq. (\ref{R}), must coincide with $A^{BS}(\vec{k}_{1\perp},\vec{k}_{2\perp})$, Eq. (\ref{L}).  If the LF wave function 
is found from Eq. (\ref{eq30}) and the BS amplitude is found from Eq. (\ref{3bE}), then $A^{LF}$ and $A^{BS}$ differ because of different input in the kernels of Eqs. (\ref{eq30}) and  (\ref{3b}). The comparison of $A^{LF}$ with $A^{BS}$  shows the influence of the 
many-body intermediate states on the \mbox{$k_{\perp}$-dependence} of the amplitude $A^{BS}$.

\section{The two lowest-lying levels}\label{levels}

In this section we  present the numerical results for the ground and first excited states. Both LF and BS equations are solved by means of spline decomposition and the results are presented within the  convergence $\lesssim 3 \%$, which is enough for our purposes.

\begin{figure}[!hbt]
\centering
\epsfig{figure=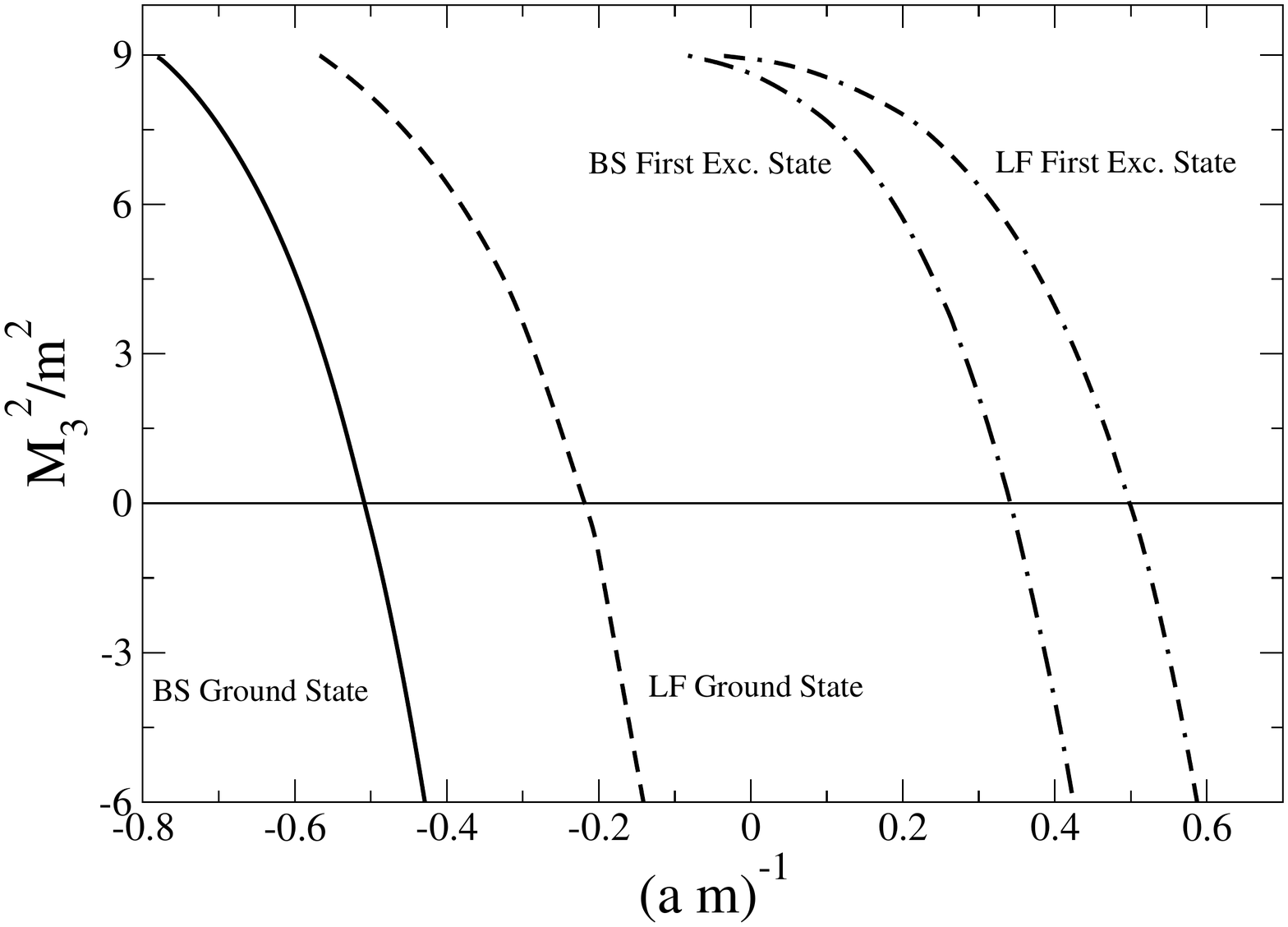,width=10cm}
\vspace{-1cm}
\caption{The value $M_3^2$ vs. the inverse scattering length $(a\,m)^{-1}$.  BS ground state (solid curve); LF ground state (dashed curve);  BS first excited state (dashed-dotted) and  LF first excited state (double-dash-dotted). }
\label{Fig4} 
\end{figure}

We expect that the spectra of both equations are rather rich. 
However, as we said, we restrict ourselves to two low-lying states. 
The LF equation (\ref{eq30}) determines the value $M_3^2$.
The situation with the BS equation (\ref{3bE}) is the same.  At a  first glance, the BS equation  determines $M_3$ in 
the first degree. However, the change of the sign $M_3\to -M_3$ is equivalent to the complex conjugation, which does not 
change the real eigenvalues. Hence, Eq. (\ref{3bE}) also determines $M_3^2$. Though $M_3^2$  originally appears as squared, when this parameter is found from the equations, it can have any sign. 
The relativistic effects eliminate the Thomas collapse,  i.e., they do not allow the eigenvalues $M_3^2$ to decrease down to $-\infty$, 
though they do not prevent the value of $M_3^2$ from being negative for strong enough two-body interaction. 
It turns out that ``strong enough" is already the interaction forming a two-body state with the binding energy close to zero -- it provides negative $M_3^2$ for the ground state. However, when we further weaken the two-body interaction  (the scattering length becomes negative and then $|a|\to 0$), the ground state value of $M_3^2$ becomes positive and then $M_3\to 3m$ ($B_3= 3m-M_3\to 0$), i.e., the three-body bound states disappear. The plot of $M_3^2$ vs. the inverse scattering length $(a\,m)^{-1}$ is shown in Fig. \ref{Fig4}.  Note that in the previous papers \cite{tobias1,ck3b} just the ``LF-excited state" was studied. Our present calculations confirm the values $M_3^2$ vs. $M_2$ found in \cite{ck3b}.

\begin{table}[!]
\begin{center}
\begin{tabular}{|c|c|c|c|c|}
\hline 
 & \multicolumn{4}{|c|}{Inverse scattering length $(a\,m)^{-1}$}\\
\cline{2-5}
$M_3^2$& \multicolumn{2}{|c|}{ground state}& \multicolumn{2}{|c|}{excited state}\\
\cline{2-5}
 &BS&LF&BS&LF\\
\hline
$9m^2$& $-0.78$ & $-0.57$ & $-0.08$ & $-0.04$ \\
\hline
$0$& $-0.51$ & $-0.21$ & $\phantom{-}0.34$ & $\phantom{-}0.50$ \\
\hline
\end{tabular}
\end{center}
\caption{The values of the inverse scattering length $(a\,m)^{-1}$ for which the curves in Fig. \protect{\ref{Fig4}} cross $M_3^2=9m^2$ (that is, $B_3=3m-M_3=0$) and $M_3^2=0$. Values presented within the used convergence, as discussed in the text.}
\label{tab1}
\end{table}

Fig. \ref{Fig4} shows that the three-body mass $M_3^2$ found in the BS approach is always smaller than $M_3^2$ found in the LF one. This means that  the three-body forces discussed in Sec. \ref{lfeq} are attractive and  strong. This conclusion coincides with the result found in Refs. \cite{mc_2000,km08} for the OBE kernel. 
 The dimensionless values $(a\,m)^{-1}$ for which  the values $M_3^2$ cross zero and cross $9m^2$ (when the three-body binding energy $B_3=3m-M_3$ crosses zero) are given in the Table \ref{tab1}. The positive inverse scattering lengths $(a\,m)^{-1} \approx 0.34$ (BS) and $(a\,m)^{-1} \approx 0.50$ (LF), for which $M_3^2$ for excited state crosses zero, correspond, according to Eq. (\ref{a}),  to 
the two-body binding energies $B_2 \approx 0.194m$ and $B_2 \approx 0.582m$, respectively. When $B_2=2m-M_{2}\to 0$, the ground state values  
are $M_3^2 \approx -94 \, m^2$ for the BS equation and $M_3^2=-18 \, m^2$ for the LF one. They are extremely over-bounded.  The corresponding excited state values (when $B_2=0$) are $B_3 \approx 0.066m$  for the BS equation and  $B_3 \approx 0.013m$ for the LF one. The latter value is close to one computed in \cite{ck3b}.

\begin{figure}[thb]
\centering
\epsfig{figure=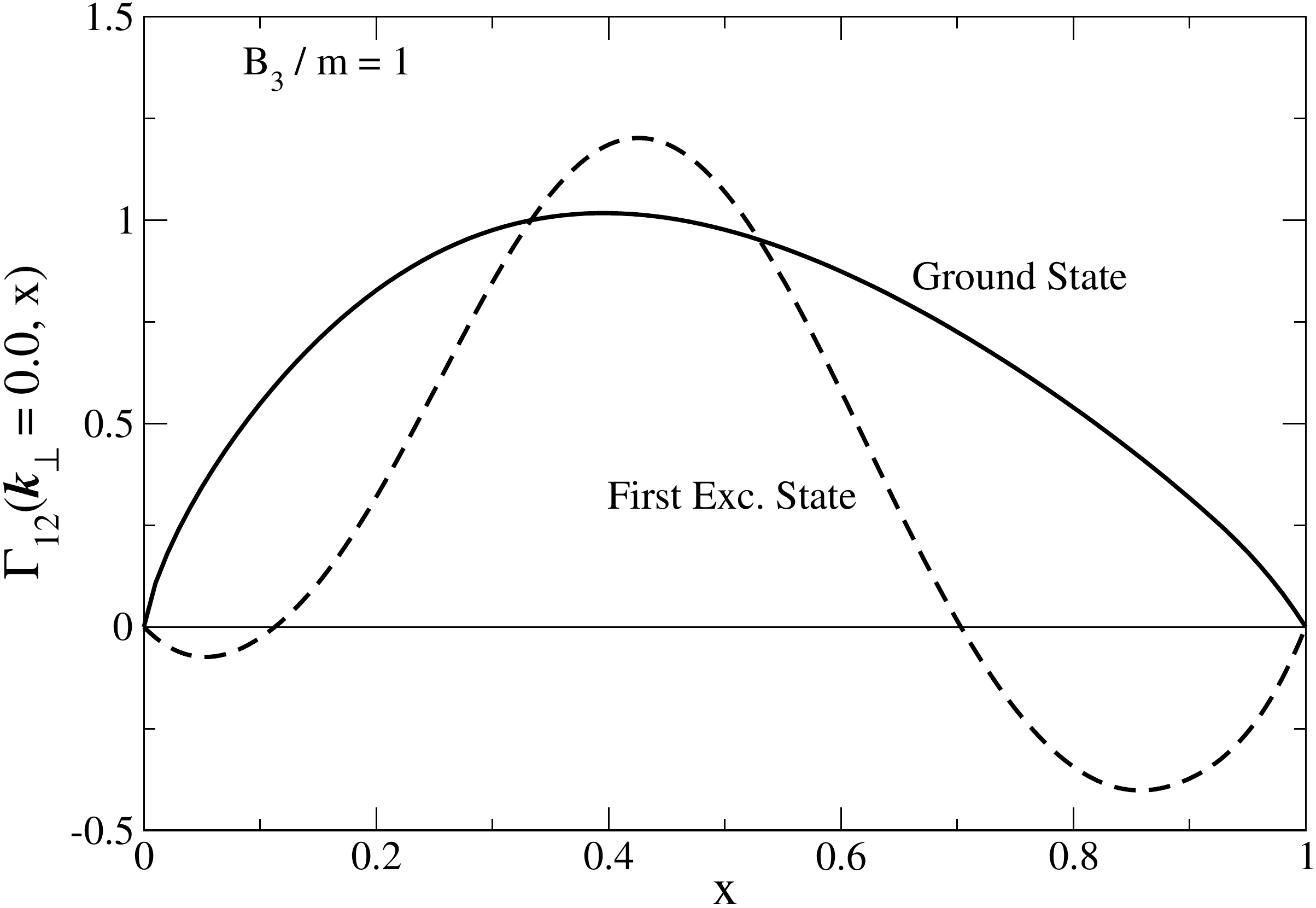,width=7.5cm}
\quad
\epsfig{figure=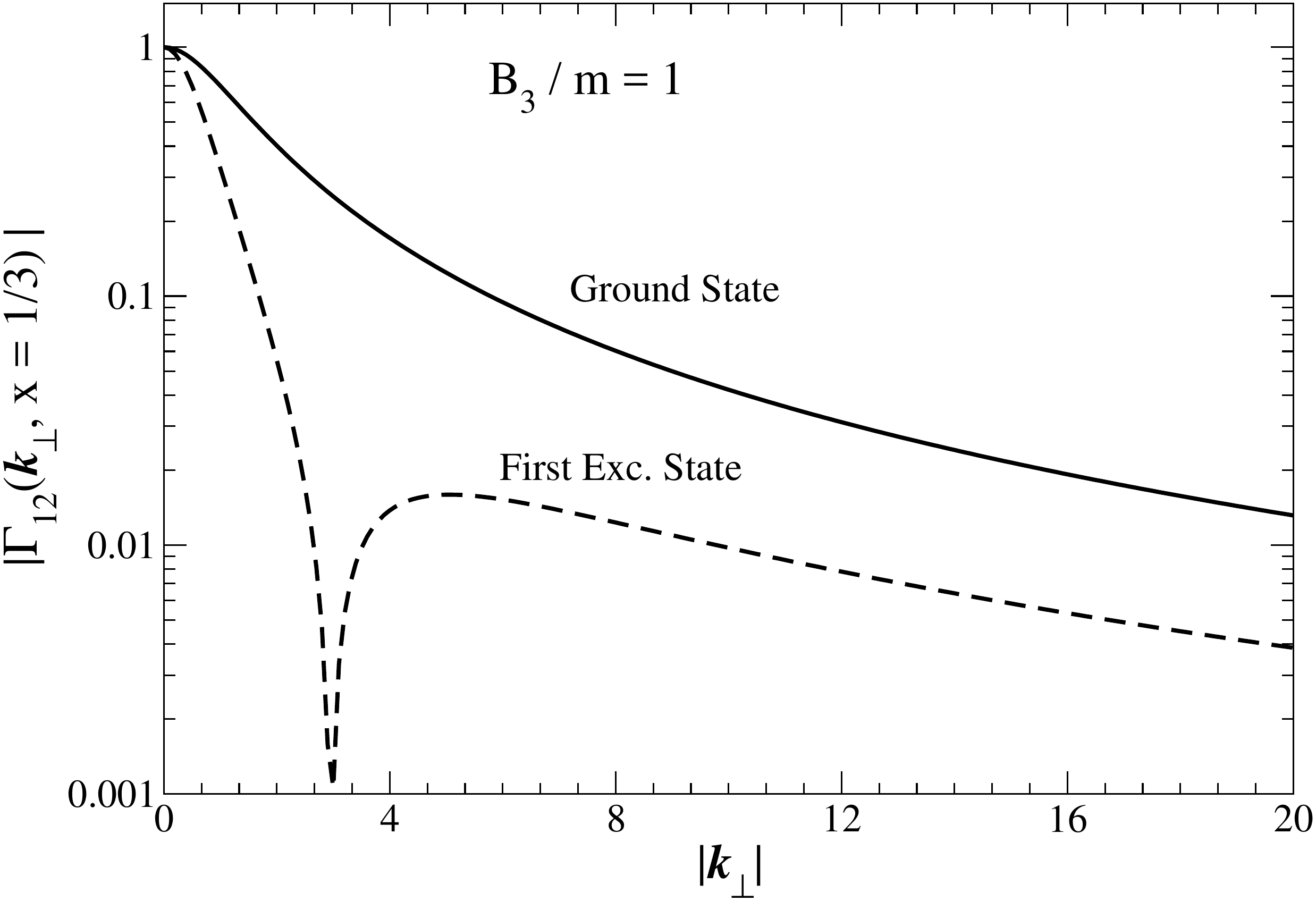,width=7.5cm}
\caption{The vertex function $\Gamma(k_{\perp}=0,x)$, satisfying the LF equation (\ref{eq30}), vs. $x$ (left panel);
and  $\Gamma(k_{\perp},x=1/3)$ vs. $k_{\perp}$ (right panel).  In both panels we present the
  ground state with $B_3=m$ (solid  curve); and the excited state (dashed curve), also with $B_3=m$, but for different $(a\,m)^{-1}$, both given in the text.}\label{Fig6} 
\end{figure}

\section{Light-front vertex function and Bethe-Salpeter amplitude}\label{LFBS}

In order to study how the binding energy impacts the behavior of the solution, we vary the two-body parameters  to obtain, in the LF framework,  the binding energy $B_3=m$, first, for the ground state (in this case: $(a\,m)^{-1}=-0.31$), then, for the excited state (in this case: $(a\,m)^{-1}=0.4\to B_2=0.297m$). These two solutions $\Gamma(k_{\perp},x)$ of  the LF equation (\ref{eq30}), corresponding to the same binding energy, but to different states (ground and excited ones) and normalized by  $\Gamma(0,1/3)=1$,
are compared in Fig. \ref{Fig6}. Though, in general, they considerably differ from each other, the functions  $\Gamma(k_{\perp},x=1/3)$ vs. $k_{\perp}$ for the ground and excited states have the same asymptotic decrease, though with different coefficients: $\Gamma$ for the excited state is ten times smaller than  for the ground state.

\begin{figure}[thb]
\centering
\epsfig{figure=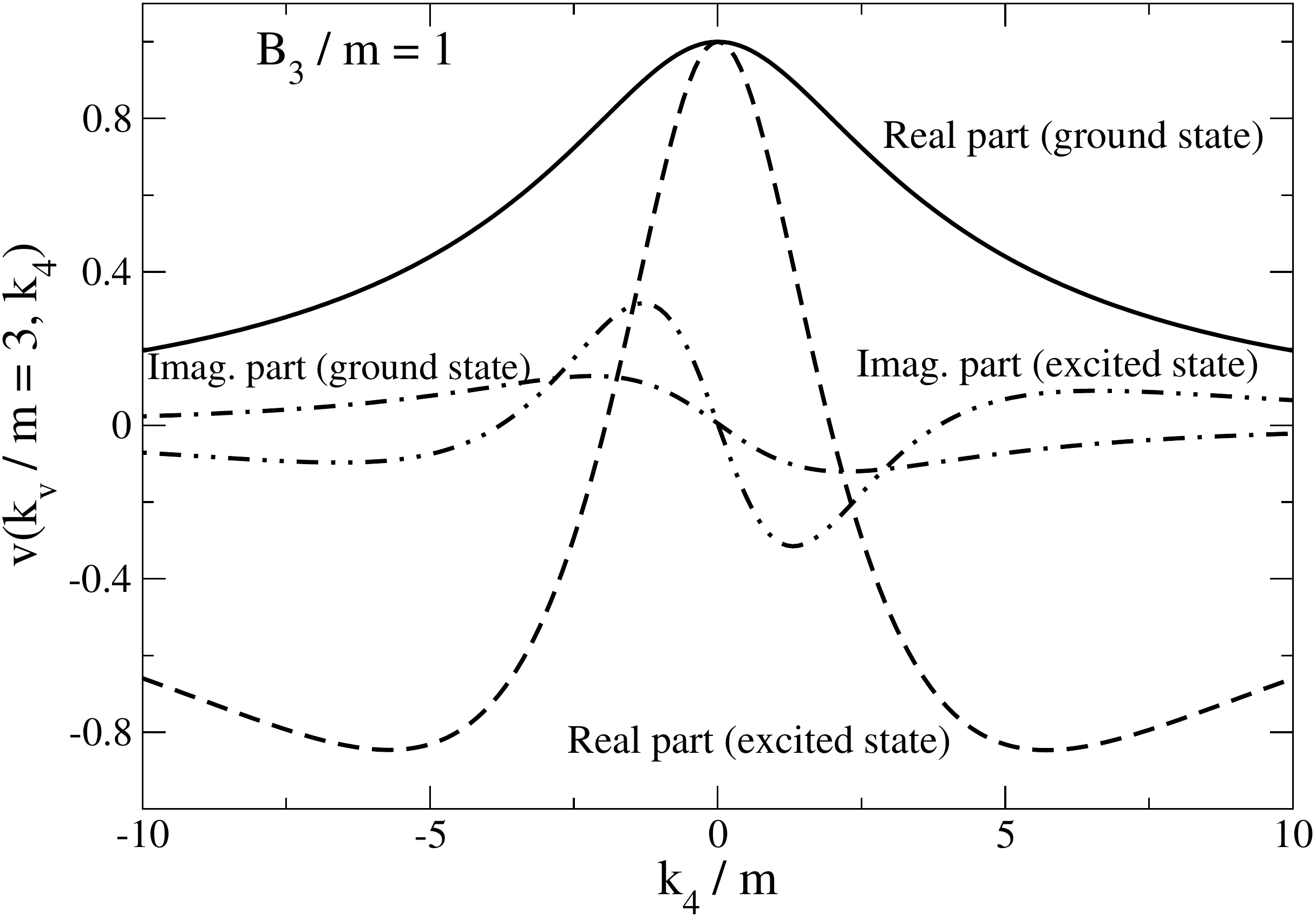,width=7.5cm}
\quad
\epsfig{figure=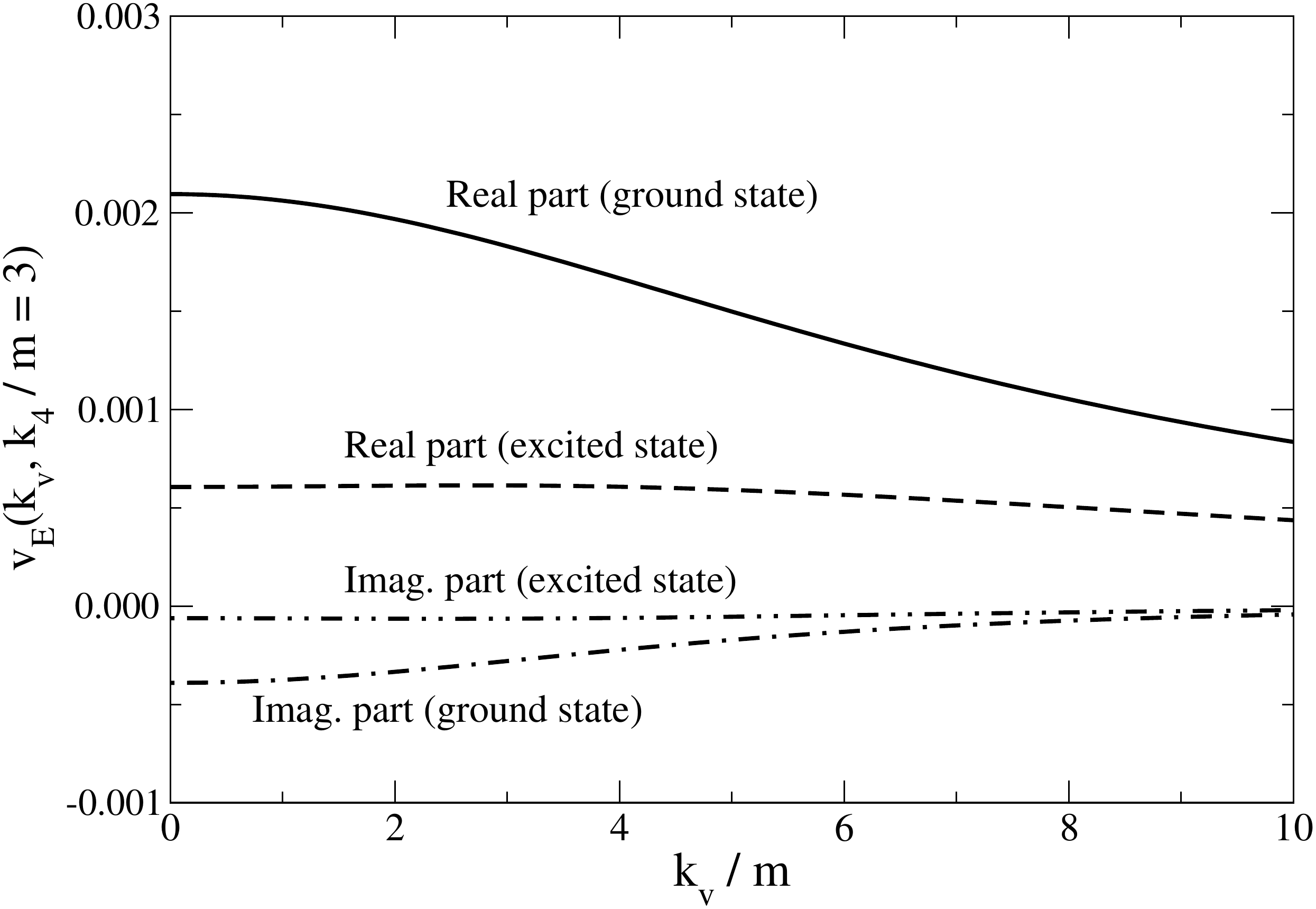,width=7.5cm}
\caption{The BS amplitude $v_E(k_4,k_v/m=3)$ vs. $k_4$ normalized to ${\rm Re}[v_E(k_4=0,k_v/m=3)]=1$. Solid (ground state) and dashed (exited state) curves are ${\rm Re}[v_E(k_4,k_v/m=3)]$ (left panel) and ${\rm Re}[v_E(k_4/m=3,k_v)]$ (right panel); dot-dashed (ground state) and dot-dot-dashed (exited state) curves are ${\rm  Im}[v_E(k_4,k_v/m=3)]$ (left panel) and ${\rm Im}[v_E(k_4/m=3,k_v)]$ (right panel).}
\label{Fig7} 
\end{figure}

The asymptotic behavior of $\Gamma(k_{\perp},x)$ follows from Eq. (\ref{eq30}). Up to the logarithmic correction resulting from $F(M_{12})$, the asymptotic  $k_{\perp}$-dependence is provided by the factor $(M_0^2-M_3^2)\sim k_{\perp}^2$ that gives $\Gamma(k_{\perp},x)\sim c/k_{\perp}^2$, which is close to the asymptotic form of both curves shown in the right panel of Fig. \ref{Fig6}. Whereas, $\Gamma$ in the the non-asymptotic domain and the  factor $c$ are determined by the integral in l.h.  side of Eq. (\ref{eq30}) which is sensitive to the details of $\Gamma(k_{\perp},x)$. Therefore they strongly depend on the state.

The solutions $v_E(k_4,k_v)$ of the Euclidean BS equation (\ref{3bE}) for $B_3\,=\,m$ 
[$(a\,m)^{-1} \approx -0.57$ for the ground state and $(a\,m)^{-1}=0.25\to B_2=0.093\,m$ for the excited state] are shown in Fig. \ref{Fig7}. Note that ${\rm Re}[v_E(k_4,k_v=\mathrm{const})]$ vs. $k_4$ is symmetric  relative to $k_4\to -k_4$ and  ${\rm Im}[v_E(k_4,k_v=\mathrm{const})]$ is antisymmetric,
in accordance with Eq. (\ref{sym}).

\begin{figure}[thb]
\centering
 \includegraphics[scale=0.32,angle=0]{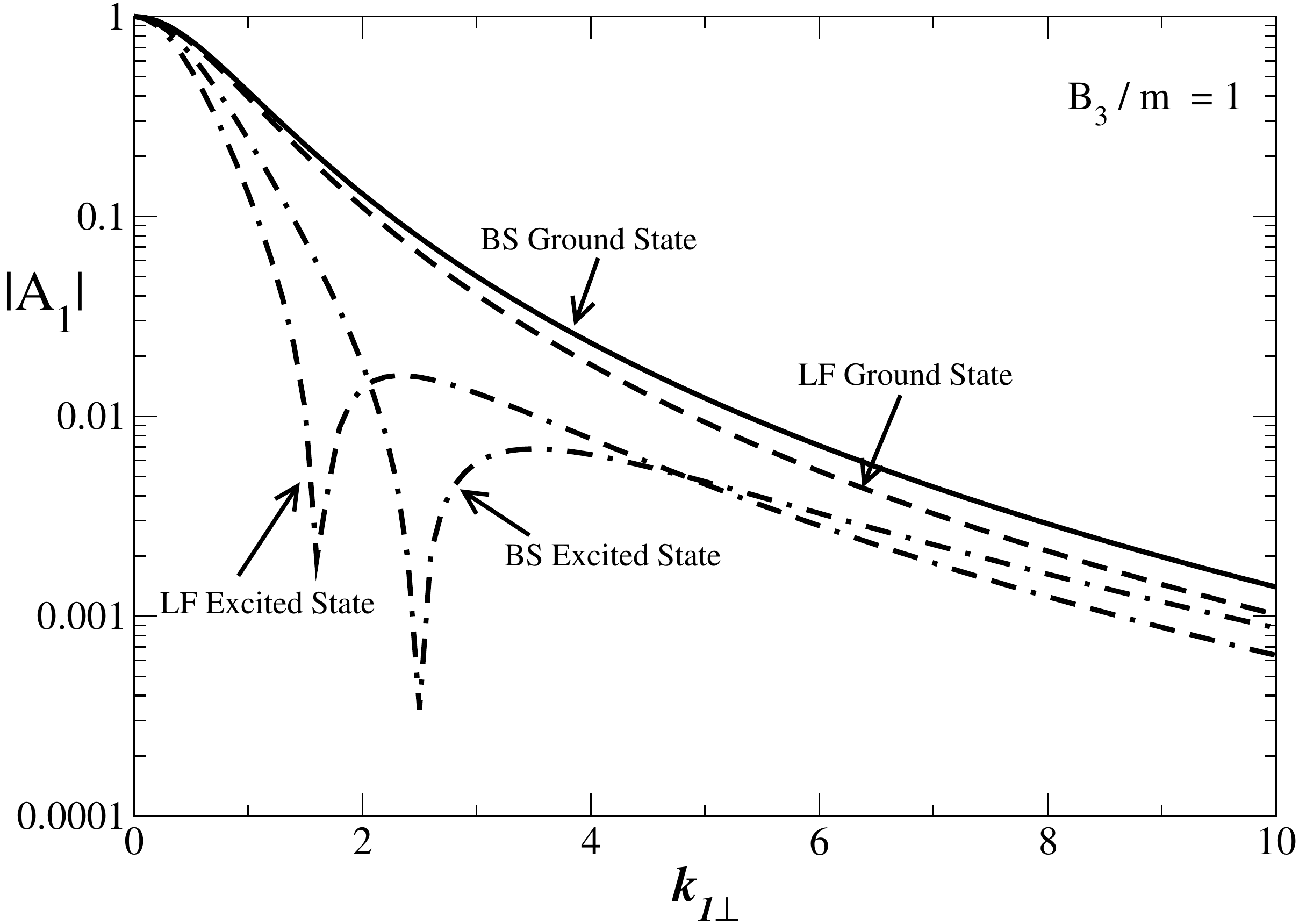}
\caption{$k_{\perp}$- dependences of the Faddeev components of the 
LF and  Euclidean amplitudes for the same binding energy, $B_3/m=1$. 
The solid  and dot-dashed
curves  are the BS calculations (Eq. (\ref{L})) for the ground and first excited state, respectively. 
The dashed  and dash-dash-dotted  curves are the LF calculations (Eq. (\ref{R})), for the ground and first excited state, respectively. }\label{Fig8} 
\end{figure}

The comparison of the $k_{\perp}$-dependences of the LF and BS amplitudes is shown in Fig. \ref{Fig8}. Though the full amplitudes are given by sums of three Faddeev components (Eq. (\ref{R}) in the LF approach and Eq. (\ref{L}) in the BS approach), we present the contributions of one component only, i.e.,  $A^{LF}_{1}$, Eq. (\ref{R}), in comparison  to
$A^{BS}_{1}$, Eq. (\ref{L}), each of them depends on $\vec{k}_{1\perp}$ and $\vec{k}_{2\perp}$. We put $\vec{k}_{2\perp}=0$, normalize both $A_{1}$  to 1 at $k_{1\perp}=0$ and compare their $k_{1\perp}$ dependencies. The calculations were carried out for $B_3\,=\,m$ in both approaches. 
The node structure is clearly visible in the figure. This is important since the number of nodes is a way of characterizing states and the ground state has no node, while  the first excited state presents one.

One can see in Fig.\ref{Fig8} that for the same three-body binding energy, the BS approach results in 
a wider distribution than the LF one. This reflects the effect 
coming from the three-body  graphs that are not considered in the LF truncated equation. If we compare the $k_{\perp}$- dependences obtained from Minkowski and Euclidean BS equations we should obtain the same result, as shown in~\cite{cristian} for the two-body case. In any given approach (BS or LF), the   large momentum behavior  of the excited state is the same as for the ground one, though BS and LF asymptotics look slightly different from each other.

\section{Conclusion}\label{concl}
We have found the ground and first excited state solutions for the three-boson system with zero-range interaction in the framework of two relativistic approaches: Bethe-Salpeter equation in the Euclidean space and light-front dynamics. In the BS framework, the solution was found for the first time. Our input is the two-body scattering length (or  binding energy), the output is the three-body binding energies, the  light-front wave functions and  Bethe-Salpeter amplitudes. 

We confirmed the value of binding energy found previously \cite{ck3b} in the LF 
framework. In addition, we found that the calculations \cite{tobias1,ck3b} dealt  with 
the first excited state, though for the two-body interaction which allows the two-body 
bound state (used in \cite{tobias1,ck3b}), there is a  three-body ground state but with 
non-physical negative squared mass, $M_3^2<0$. This solution formally exists, but not as 
a physical state. The negative (though finite) $M_3^2$ can be interpreted as collapse of a relativistic system. 
However, for  a two-body interaction characterized by negative scattering length (i.e.~no two-body 
bound state), the aforementioned three-body state becomes physical, i.e.~having positive 
$M_3^2$. We get a strongly bound Borromean system for the negative scattering length, 
that is rather curious. Another way to avoid the negative $M_3^2$ is to introduce a 
cutoff.
We expect that a  cutoff can also weaken the two-body interaction and make  
$M_3^2$ positive.  By a further decrease of the two-body interaction the three-body 
binding energy tends to zero so that the three-boson  system becomes unbound.

We have also found that
in spite of the same zero-range interaction, the dynamical contents in both approaches 
-- BS and LF -- are different. Relative to the LF dynamics, the BS approach implicitly 
takes into account the antiparticles and the many-body intermediate states which 
generate the effective three-body forces of the relativistic origin, like it happens for the OBE 
kernel \cite{km08}, but with smaller diversity of the graphs contributing  to three-body 
forces. However, their net effects is the same -  the increase of the effective 
attraction and, consequently,  the  binding energy in the BS framework with respect to 
the LF one. At the same time, the fully relativistic effects in both 
frameworks are the effective repulsion, eliminating the Thomas collapse 
\cite{thomas} in a three-boson system.  This was found earlier in the LF approach 
\cite{tobias1,ck3b}. In the present paper this is confirmed also in the BS approach. 

A comparison of the LF wave function with the Euclidean BS amplitude cannot  be done directly, since these quantities have different nature and physical meaning. However, as it is shown in Sec. \ref{trans},  the integrals calculated from both quantities,
either over $x$, or over $k^-$ and $k_z$,  represent  one and the same amplitude depending on
transverse momenta (provided, the underlying dynamics is the same). At the same time, 
the contributions from three-body forces discussed in Sec. \ref{lfeq}, that makes 
different the binding energies, also affects the $k_{\perp}$-amplitudes. We compared 
these amplitudes for the same binding energy $B_3$ of the three-body system and we found 
that  in the BS approach the  $k_{\perp}$-distributions are somewhat wider 
than in the LF one. 

The solutions and observations found in this work deepen our understanding of the role 
of relativistic effects in  three-body systems. This research can be 
generalized to systems with non-equal masses  \cite{SuiPRD02}, which naturally may have 
a richer spectrum.

{\it Aсknowledgements.} We thank the support from Conselho Nacional de Desenvolvimento Cient\'ifico e Tecnol\'ogico (CNPq) and 
Coordena\c c\~ao de Aperfei\c coamento de Pessoal de N\'ivel Superior 
(CAPES) of Brazil. J.H.A.N. acknowledges the support of the grant \#2014/19094-8 and V.A.K. of the grant \#2015/22701-6 from
 Funda\c c\~aode Amparo \`a Pesquisa do Estado de S\~ao Paulo (FAPESP).  V.A.K.  is also sincerely grateful to group of theoretical nuclear physics of ITA, S\~{a}o Jos\'e dos Campos, Brazil, for kind hospitality during his visit.

\end{document}